\documentclass[a4paper]{article}
\usepackage{geometry}
\usepackage{amsmath}
\usepackage{CJK}
\usepackage{chemarrow}
\usepackage{setspace}
\usepackage{array}
\usepackage{graphicx}
\usepackage{subfigure}

\begin{document}

\begin{spacing}{1.5}
\setlength{\parindent}{2em}
 \numberwithin{equation}{section}

\title{Asymptotic analysis of the lattice Boltzmann method for generalized Newtonian fluid
flows}
\author
{Zai-Bao Yang\footnote{Zhou Pei-Yuan Center for Appl. Math., Tsinghua Univ.,
Beijing 100084, China;
Email: yang-zb11@mails.tsinghua.edu.cn},
~Wen-An Yong\footnote{Zhou Pei-Yuan Center for Appl. Math., Tsinghua Univ.,
Beijing 100084, China; Email:
Email: wayong@tsinghua.edu.cn}}
\date{}
\maketitle

\begin{abstract}

In this article, we present a detailed asymptotic analysis of the lattice Boltzmann method
with two different collision mechanisms of BGK-type on the D2Q9-lattice for
generalized Newtonian fluids. Unlike that based on the Chapman-Enskog expansion leading
to the compressible Navier-Stokes equations, our analysis gives the
incompressible ones directly and exposes certain important features of the lattice Boltzmann solutions.
Moreover, our analysis provides a theoretical basis
for using the iteration to compute the rate-of-strain tensor, which makes sense
specially for generalized Newtonian fluids.
As a by-product, a seemingly new structural condition on the generalized Newtonian fluids
is singled out. This condition reads as ``the magnitude of the stress tensor increases
with increasing the shear rate". We verify this condition
for all the existing constitutive relations which are known to us.
In addition, it it straightforward to extend our analysis to MRT models
or to three-dimensional lattices. \\

\end{abstract}

\section{Introduction}

During the last two decades, the lattice Boltzmann method (LBM) has been developed into
an effective and viable tool for simulating various fluid flow problems. It has
been proved to be quite successful in simulating complex Newtonian fluid flows such as
turbulent flows, micro-flows, multi-phase and multi-component flows,
particulate suspensions, and interfacial dynamics.
We refer to
\cite{LKS, YMLS, CD} for a comprehensive account of the method and its applications.
Moreover, the potential of LBM in simulating flows of generalized Newtonian fluids,
for which the dynamic shear viscosity depends on the shear rate \cite{BAH}, was
shown by Aharonov and Rothman \cite{E.A} as early as 1993. In recent years,
this potential has attracted much attention
\cite{Gabbanelli, Boyd, Yoshino, Psihogios, Boyd1, Wang1, Tang, Tang1, Wang, Guo, Oh}.

There are at least two kinds of LB-BGK models for generalized Newtonian fluids.
In the first model, the relaxation time is not a constant anymore but depends on the shear rate.
The non-Newtonian effects are embedded in the LBM through a dynamic change of the
relaxation time. The second model has a constant relaxation time but its equilibrium
distribution contains the shear rate.


The original goal of this paper is to extend the asymptotic analysis of LBM \cite{JY, JKL}
for classical Newtonian fluids to generalized ones. The analysis is based on the diffusive,
instead of convective, scaling developed in \cite{DEL} for the Boltzmann equation
and in \cite{JY} for LB equations. It turns out that the diffusive scaling is the most
natural choice if the LBE is viewed as a numerical solver of the incompressible Navier-Stokes
equations. Unlike that based on the Chapman-Enskog expansion leading to the compressible
Navier-Stokes equations (see, e.g. \cite{Wang1}), our analysis uses the Hilbert expansion and gives the incompressible
equations directly. Meanwhile, this analysis exposes certain important features of the LB solutions.
In particular, it provides a theoretical basis for using the Richardson extrapolation
technique to improve the accuracy up to higher orders and for using the iteration \cite{Boyd, Tang, Guo} to compute
the rate-of-strain tensor. The latter makes sense specially for generalized Newtonian fluids.

To achieve the above goal, we realize that a proper introduction of
the lattice spacing in the equilibrium distribution functions are
indispensable to recover the correct macroscopic equations. Moreover,
we have to carefully dispose many possibly non-zero terms due to the
dependence on the shear rate, which vanish for classical Newtonian fluids.

As a by-product of our analysis, a seemingly new structural condition on the generalized Newtonian fluids
is singled out. This condition reads as ``the magnitude of the stress tensor increases
with increasing the shear rate" (magnitude of the rate-of-strain tensor). We verify this condition
for all the existing constitutive relations which are known to us. By inductive reasoning,
we suggest to take this structural condition as a criterion in constructing further constitutive
relations for generalized Newtonian fluids.

In addition, let us mention that our analysis can be easily extended to the multiple-relaxation-time (MRT) models \cite{dH, LL, dhumieres02} or to three-dimensional lattices, although it is presented only for BGK models on the two-dimensional lattice D2Q9 \cite{QDL}.

This paper is organized as follows. In Section 2 we introduce the
LBM together with the D2Q9 lattice and two collision terms for
generalized Newtonian fluids. The asymptotic analysis is outlined
in Section 3. In Section 4, we verify the structural condition for the existing
constitutive relations known to us. Several
conclusions are summarized in Section 5. Finally, an appendix
is devoted to the technical details skipped in Section 3.

\section{Lattice Boltzmann Method}

The general form of the lattice Boltzmann method is
\begin{equation}\label{21}
f_i(x + c_ih, t + \delta t) - f_i(x, t) = \Omega_i(f_1, f_2,
\cdots, f_N)(x, t)
\end{equation}
for $i =0, 1, 2, \cdots, N$. Here $N$ is a given integer,
$f_i=f_i(x, t)$ is the $i$-th density distribution function of particles at the
space-time point $(x, t)$,
$c_i$ is the $i$-th given velocity, $h$ is the lattice
spacing, $\delta t$ is the time step, and $\Omega_i= \Omega_i(f_1,
f_2, \cdots, f_N)$ is the $i$-th given collision term. Motivated by
the diffusive scaling analyzed in \cite{DEL, JY}, we take
$$
\delta t= h^2
$$
in what follows.

For the sake of definiteness, we take the D2Q9 lattice throughout
this paper. Namely, our problem is two-dimensional, $N=8$ and
$$
c_{i}=
\begin{cases}
(0,0) & i=0\\
(1,0),(0,1),(-1,0),(0,-1) & i=1,2,3,4\\
(1,1),(-1,1),(-1,-1),(1,-1) & i=5,6,7,8.
\end{cases}
$$
For $i\in\{0, 1,2, \cdots, 8\}$, we define
\begin{equation}\label{22}
\bar{i}=
\begin{cases}
0 & i=0\\
3,4,1,2 & i=1,2,3,4\\
7,8,5,6 & i=5,6,7,8.
\end{cases}
\end{equation}
It is clear that $c_{i}=-c_{\bar i}.$ In this sense, $c_i$ is said
to be {\it odd}.

We will discuss two different collision mechanisms of BGK-type for generalized
Newtonian fluids \cite{BAH}. Unlike for classical Newtonian fluids, the dynamic shear
viscosity for the generalized ones is not a constant but depends on
the gradient of the fluid velocity $v$. Precisely, the viscosity
$\mu=\mu(I)$ is a non-negative function of $I = tr(S^2)$ with $S$
the rate-of-strain tensor or rate-of-deformation tensor
\begin{equation}\label{24}
S = \nabla v + (\nabla v)^t
\end{equation}
where the superscript $t$ indicates the transpose. The corresponding
stress tensor $T$ is given as
$$
T=\mu(I)S.
$$

The first collision term reads as
\begin{equation}\label{25}
\Omega_{i}=\frac{1}{\tau(S/h)}\big(f_{i}^{eq}(\rho, v)-f_{i}\big).
\end{equation}
Here the relaxation time is taken as
$$
\tau(S)=3\mu(I) + \dfrac{1}{2},
$$
while the equilibrium distribution is quite standard:
\begin{equation}
f_{i}^{eq}=f_{i}^{eq}(\rho, v)=w_{i}[\rho+3c_{i}\cdot
v+\dfrac{9}{2}(c_{i}\cdot v)^2-\dfrac{3}{2}v\cdot v]
\end{equation}
($\cdot$ indicates the inner product) with
\begin{equation}\label{27}
\rho=\sum_{i=0}^8 f_{i}, \qquad v=\sum_{i=0}^8 c_{i}f_{i}
\end{equation}
and weighted coefficients
\begin{equation}\label{28}
w_{i}=\dfrac{1}{36}
\begin{cases}
16 & i=0\\
4 & i=1,2,3,4\\
1 & i=5,6,7,8.
\end{cases}
\end{equation}
Note that $w_i=w_{\bar i}$, that is, $w_i$ is {\it even}.

For convenience, we decompose the equilibrium distribution above as
\begin{equation}
f_i^{eq} = f_{iL}(\rho, v) + f_{iQ}(v, v)
\end{equation}
with
\begin{equation}
\begin{array}{l}
f_{iL}(\rho,v):=w_{i}(\rho+3c_i\cdot v),\\[4mm]
f_{iQ}(u,v):=w_i[\dfrac{9}{2}(c_i\cdot u)(c_i\cdot
v)-\dfrac{3}{2}u\cdot v].
\end{array}
\end{equation}
Since $c_i$ is odd and $w_i$ is even, it is easy to see that
\begin{equation*}
\begin{array}{rl}
\sum_i f_{iL}(\rho, v)\equiv \rho, \qquad \sum_if_{iQ}(u, v)\equiv
0,\\[4mm]
\sum_ic_if_{iL}(\rho, v)\equiv v, \qquad \sum_ic_if_{iQ}(u, v)\equiv
0.
\end{array}
\end{equation*}

The second collision term is
\begin{equation}\label{211}
\Omega_{i}=\frac{1}{\tau}\big(f_{i}^{eq}(\rho, v; S, h) - f_{i}\big)
\end{equation}
with a constant relaxation time $\tau$. The equilibrium distribution
is \cite{Yoshino, Wang1}
\begin{equation}\label{212}
f_{i}^{eq}(\rho, v; S, h)=f_{iL}(\rho, v) + f_{iQ}(v, v) +
w_ihA(\frac{S}{h})S:c_i^tc_i
\end{equation}
with scalar function
$$
A(S)=\frac{3}{2}(\tau-\frac{1}{2})-\frac{9}{2}\mu(trS^2).
$$
In \eqref{212} $S:c_i^tc_i$ is the standard contraction of two
symmetric tensors $S$ and $c_i^tc_i$.

It is remarkable that the lattice spacing $h$ has been introduced in
the two collision terms \eqref{25} and \eqref{211}, which seems new.


\section{Asymptotic Analysis}

Motivated by Strang \cite{St}, we notice that the LB solution
$f_i=f_{i}(x, t; h)$ depends on the lattice spacing $h$ which is
small. Thus, we will seek an expansion of the form
\begin{equation}\label{31}
f_{i}(x, t; h)\sim \sum_{n\geq 0}h^n f_{i}^{(n)}(x,t).
\end{equation}
Referring to this expansion, \eqref{27} and \eqref{24}, we introduce
\begin{equation}\label{32}
\rho^{(n)}:=\sum_{i=0}^8 f_{i}^{(n)},\qquad v^{(n)}:=\sum_{i=0}^8
c_if_{i}^{(n)}, \qquad S^{(n)}:=(\nabla v^{(n)})+(\nabla
v^{(n)})^{T}
\end{equation}
and
\begin{equation}\label{33}
\rho_h\sim\sum_{n\geq 0}h^n\rho^{(n)},\qquad v_h\sim\sum_{n\geq 0}h^n
v^{(n)},\qquad S_h\sim\sum_{n\geq0}h^n S^{(n)}.
\end{equation}


Take $f_i^{(0)}=w_i$ as the leading term in \eqref{31}. It follows
clearly from \eqref{32} together with the even/odd properties of
$w_i$ and $c_i$ that
$$
\rho^{(0)}=1, \qquad v^{(0)}=0, \qquad S^{(0)}=0.
$$
By using the Taylor expansion, we have
\begin{equation}\label{34}
f_{i}^{(n)}(x+c_{i}h,t+h^2)-f_{i}^{(n)}(x,t)\sim\sum_{l>0}\dfrac{(h^2\partial_t+h c_i\cdot \nabla)^l}{l!}f_{i}^{(n)}(x,t).
\end{equation}
Since
$$
\begin{array}{rl}
& \sum\limits_{l>0}\dfrac{(h^2\partial_t+h c_i\cdot \nabla)^l}{l!}=\sum\limits_{l>0}\sum\limits_{m=0}^{l} \frac{1}{m!(l-m)!}(h^2\partial_t)^{m}(h c_i\cdot\nabla)^{l-m}\\[4mm]
=& \sum\limits_{s>0}h^s\sum\limits_{l+ 2m=s}\frac{1}{l!m!}\partial_t^m(c_i\cdot\nabla)^l\equiv \sum\limits_{s>0}h^sD_{i, s}
\end{array}
$$
with $D_{i, s}$ a differential operator, it follows from \eqref{34}
and the constancy of $f_i^{(0)}$ that
\begin{eqnarray}\label{35}
f_{i}(x+c_{i}h,t+h^2)-f_{i}(x,t)&=&\sum_{n\geq 0}\sum_{l>0}\sum_{m\leq l}h^{n+l+m}\dfrac{\partial_t^m(c_i\cdot\nabla)^{l-m}}{m!(l-m)!}f_i^{(n)}(x,t)\nonumber\\[4mm]
 &=&\sum_{n\geq 2}h^n\sum_{s=1}^{n-1}D_{i,s}f_{i}^{(n-s)}(x,t).
\end{eqnarray}

For the first collision \eqref{25}, we expand
\begin{equation*}
\begin{split}
\tau(S_h/h)&=\tau(S^{(1)}+h S^{(2)}+\cdot\cdot\cdot)
\sim\sum_{n\geq 0}h^nF^{(n)}.
\end{split}
\end{equation*}
It is not difficult to see that $F^{(n)}$ is determined with
$S^{(l)}$ for $l=1, 2, \cdots, n + 1$. In particular,
$F^{(0)}=\tau(S^{(1)})$.
Thus, we may rewrite the LBM \eqref{21} as
\begin{equation}\label{36}
\begin{array}{rl}
& \sum_{r\geq 0}h^rF^{(r)}\sum_{n\geq 2}h^n\sum_{s=1}^{n-1}D_{i,s}f_{i}^{(n-s)}\\[4mm]
=& -\sum_{n\geq
0}h^n[f_{i}^{(n)}-f_{iL}(\rho^{(n)},v^{(n)})]+\sum_{n\geq 1}h^n
\sum_{p+q=n}f_{iQ}(v^{(p)},v^{(q)}).
\end{array}
\end{equation}
By equating the coefficient of $h^k$ in the two sides of the last
equation and using $v^{(0)}=0$, we obtain
\begin{align}
&h^0: \quad f_i^{(0)}=f_{iL}(\rho^{(0)},v^{(0)})=f_{iL}(1, 0), \\
&h^1: \quad  f_i^{(1)}=f_{iL}(\rho^{(1)},v^{(1)}),\label{38}\\
&h^2: \quad \tau(S^{(1)})D_{i,1}f_i^{(1)}+f_i^{(2)}=f_{iL}(\rho^{(2)},v^{(2)})+
f_{iQ}(v^{(1)},v^{(1)}),\label{eq1}\\
&h^k:\quad \tau(S^{(1)})\sum_{s=1}^{k-1}D_{i,s}f_i^{(k-s)}+F^{(1)}\sum_{s=1}^{k-2}D_{i,s}f_i^{(k-1-s)}+\cdot\cdot\cdot+
F^{(k-2)}D_{i,1}f_i^{(1)}+f_i^{(k)}\nonumber\\
&=f_{iL}(\rho^{(k)},v^{(k)})+\sum_{p+q=k}f_{iQ}(v^{(p)},v^{(q)})\label{310}
\end{align}
for $k\geq 3$.
With this hierarchy of equations, the expansion coefficient
$f_i^{(k)}$ can be uniquely determined in terms of $(\rho^{(l)},
v^{(l)})$ for $l=1, 2, \cdots, k$. In contrast to the case
\cite{JY, JKL} for classical Newtonian fluids, there are more terms involving
$F^{(j)}$ with $j\geq 1$ in \eqref{310}.

In Appendix, we will show that $F^{(j)}=0$ with $j$ odd and $(\rho^{(l)}, v^{(l)})$ can be
inductively obtained by solving a hierarchy of quasilinear or linear
partial differential equations. In particular, we can show that
$\rho^{(1)}\equiv0$ and $(\rho^{(2)}, v^{(1)})$ satisfies the
following equations
\begin{equation}\label{311}
\begin{split}
\nabla\cdot v^{(1)}& = 0\\
\dfrac{\partial v^{(1)}}{\partial t}+\nabla
\frac{\rho^{(2)}}{3}+v^{(1)}\cdot\nabla
v^{(1)}&=\nabla\cdot[\dfrac{1}{3}\big(\tau(S^{(1)})-\dfrac{1}{2}\big)S^{(1)}].
\end{split}
\end{equation}
Namely, $v^{(1)}$ and $\frac{\rho^{(2)}}{3}$ are the respect velocity
and pressure of the generalized Newtonian fluid,
for the relaxation time is taken as $\tau(S)= 3\mu(I)+ \dfrac{1}{2}.$
Furthermore, we have
\newtheorem{Theorem}{Theorem}
\begin{Theorem}\label{eq13}
Assume $\rho^{(2k+1)}\mid_{t=0}=0$ and $v^{(2k)}\mid_{t=0}=0$ for
$k=0,1,2,\cdot\cdot\cdot$, and the viscosity $\mu=\mu(I)$ satisfies
$$
\mu(I)+2\min\big\{\frac{d\mu(I)}{dI}, 0\big\}I\geq 0.
$$
Then, for periodic boundary-value problems, the expansion coefficients possess the following nice property
$$
f_i^{(k)}= (-1)^kf_{\bar{i}}^{(k)}.
$$
\end{Theorem}

\noindent {\bf Remark.}\quad
{\it In the next section, the structural condition above
$$
\mu(I)+2\min\big\{\frac{d\mu(I)}{dI}, 0\big\}I\geq 0
$$
is shown to be equivalent to the statement that the magnitude of the stress tensor increases
with increasing the shear rate.}

This theorem and the equations in \eqref{311} are both valid for the second collision mechanism \eqref{211}
together with \eqref{212}. The proof is given also in Appendix, where the expansion of $\tau(S_h/h)$
is replaced by
\begin{equation*}
A(S_h/h)\sim\sum_{n\geq0}h^nA^{(n)}(S^{(1)},\cdots,S^{(n+1)})
\end{equation*}
and the analogue of \eqref{36} reads as
\begin{equation}\label{eq3'}
\begin{split}
\tau \sum_{n\geq 2}h^n\sum_{s=1}^{n-1}D_{i,s}f_{i}^{(n-s)}=
&-\sum_{n\geq 0}h^n[f_{i}^{(n)}-f_{iL}(\rho^{(n)},v^{(n)})]+
\sum_{n\geq 1}h^n \sum_{p+q=n}f_{iQ}(v^{(p)},v^{(q)})\\
&+\sum_{n\geq0}h^{n+1}\sum_{l+m=n}w_iA^{(l)}S^{(m)}:c_{i}^Tc_{i}.
\end{split}
\end{equation}

From the theorem above and \eqref{33}, we see clearly that the
density and velocity moments of the LB solution $f_i=f_i(x, t; h)$
can be expanded as
$$
\begin{array}{l}
\rho_h:=\sum_if_i\sim\sum_{n\geq 0}h^n\rho^{(n)}=1+h^2 \rho^{(2)}+h^4 \rho^{(4)}+h^6 \rho^{(6)}+\cdots,\\[4mm]
v_h:=\sum_ic_if_i\sim\sum_{n\geq 0}h^n v^{(n)}=h v^{(1)}+h^3
v^{(3)}+h^5 v^{(5)}+\cdots .
\end{array}
$$
Thus we have
\begin{equation}\label{315}
\begin{array}{l}
\dfrac{v_h}{h} -v^{(1)} = h^2 v^{(3)} + h^4 v^{(5)} + \cdots,\\[4mm]
\dfrac{\rho_h-1}{h^2} - \rho^{(2)} = h^2 \rho^{(4)} + h^4
\rho^{(6)} + \cdots.
\end{array}
\end{equation}
These relations suggest that the rescaled LB moments
$\dfrac{v_h}{h}$ and $\dfrac{\rho_h-1}{3h^2}$ should be taken as
approximations of the velocity $v^{(1)}$ and pressure $\rho^{(2)}/3$
of the generalized Newtonian fluids with second-order accuracy in
space and first-order accuracy in time. This confirms the observation of \cite{Guo} based on numerical simulations. They also provide a basis
for using the Richardson extrapolation technique to improve the
accuracy up to higher orders.

Furthermore, we can follow \cite{YL} to deduce from the theorem that
\begin{equation*}\label{eq20}
\tau(S/h)S = \dfrac{3\sum_i(f_i^{eq}-f_i)c_i\otimes c_i}{h} +
O(h^2) .
\end{equation*}
This relation hints an alternative way to implement the LB method
with the first collision mechanism, instead of computing the shear-rate tensor
$S=\nabla v + (\nabla v)^T$ directly from $v$ with finite difference
or other methods. This issue is absent for classical Newtonian fluids,
where $\mu$ and thereby $\tau$ are constant. Indeed, we could use
the relation
$$
\Sigma = \dfrac{3\sum_{i=1}^{8}(f_i^{eq}-f_i)c_i\otimes
c_i}{h^2\tau(\Sigma)}
$$
to compute $\Sigma=S/h$ and thereby $\tau(\Sigma)$ by iteration
\cite{Boyd, Tang, Guo}. In this way, the computation of $S$ involves
the LB solution only at the current lattice point.

\section{A construction criterion}

In this section, we verify the structural precondition of Theorem
\ref{eq13}:
\begin{equation}\label{41}
\mu(I)+2I\min\{\mu'(I),0\}\geq0
\end{equation}
for some widely used constitutive relations for generalized Newtonian fluids, including the power-law model
\cite{BAH}, Carreau model \cite{Yoshino}, Carreau-Yasuda model \cite{BAH}, a smoothed Bingham model
\cite{Pap}, a smoothed Casson model \cite{Neo}, and so on. We only consider the smoothed Bingham and Casson models instead of the original ones, because the smoothness of viscosity is required in our analysis.
Notice that up to now we have treated the dynamic shear viscosity $\mu=\mu(I)$ as a function of the invariant $I=tr(S^2)$.
However, in the literature it is preferred to use the magnitude $\dot\gamma=\sqrt{I/2}$ of
the rate-of-strain tensor, instead of $I$ itself.

Considering the viscosity
$\mu=\mu(\dot{\gamma})$ as a function of $\dot\gamma$, we have
$2I\mu'(I)=\dot{\gamma}\frac{d}{d\dot{\gamma}}\mu(\dot{\gamma}).$
Thus the precondition in (\ref{41}) becomes
$$
\mu(\dot{\gamma})+\dot{\gamma}\min\{{\frac{d}{d\dot{\gamma}}\mu(\dot{\gamma})},0\}\geq0.
$$
Moreover, it is not difficult to see that the last inequality is equivalent to
\begin{equation}\label{42}
\dfrac{d}{d\dot{\gamma}}[\dot{\gamma}\mu(\dot{\gamma})]\geq0,
\end{equation}
for the viscosity $\mu(\dot{\gamma})$ is non-negative.
The last one says nothing but that $\dot{\gamma}\mu(\dot{\gamma})$ is a monotone increasing function of $\dot{\gamma}\geq 0$. Remark that the stress tensor $F=\mu(\dot{\gamma})S$ and
$$
F:F=\mu^2I=2\mu(\dot{\gamma})^2\dot\gamma^2.
$$
The above condition is just that {\it the magnitude of
the stress tensor increases with increasing the magnitude of
the rate-of-strain tensor}.

Now we turn to several concrete models. The power-law model is given in \cite{BAH} as
$$
\mu(\dot\gamma)=\mu_p\dot\gamma^{n-1},
$$
which contains two parameters $\mu_p$ and $n$. Here $\mu_p$ is the flow
consistency coefficient and $n$ is the power-law index of fluid.
According to the index $n$, the power-law fluid can be divided into three
different types. The case $n<1$ corresponds to a shear-thinning or
pseudo-plastic fluid, which is widely used in practice, whereas
$n>1$ corresponds to a shear-thickening or dilatant fluid, and $n=1$
reduces to the classical Newtonian fluid. Obviously, $\dot\gamma\mu(\dot\gamma)$ is a monotone
increasing function of $\dot\gamma\geq 0$ if $n\geq0$.

Another model for shear-shinning fluids is the Carreau model which
is preferred and used more widely in industrial applications than
the power-law model. Its viscosity is given in \cite{Yoshino} as
\begin{equation*}\label{eq4'}
\mu(\dot\gamma) = \mu_\infty+(\mu_0-\mu_\infty)[1+(\lambda\dot\gamma)^2]^{(n-1)/2} \quad\quad\text{for $0<n\leq1$} .
\end{equation*}
Here $\mu_0$ is the zero-shear-rate viscosity($\dot\gamma\rightarrow 0$), $\mu_\infty$ is the
infinity-shear-rate viscosity ($\dot\gamma\rightarrow \infty$), and $\lambda$ is a time constant.
Notice that $\mu_0>\mu_\infty$ for shear-shinning fluids. We compute
\begin{equation*}
\dfrac{d}{d\dot\gamma}[\dot\gamma\mu(\dot\gamma)] =\mu_\infty+(\mu_0-\mu_\infty)[1+\lambda^2\dot\gamma^2]^{\frac{n-3}{2}}(1+n\lambda^2\dot\gamma^2)>0
\end{equation*}
for $0<n\leq1$ and $\mu_0>\mu_\infty$. Namely, $\dot\gamma\mu(\dot\gamma)$ is a monotone
increasing function of $\dot\gamma\geq 0$.

A slightly generalization of the above model is called Carreau-Yasuda model \cite{BAH}
$$
\mu(\dot\gamma)=\mu_\infty+(\mu_0-\mu_\infty)[1+(\lambda\dot\gamma)^a]^{(n-1)/a},
$$
where the parameters have the same meaning as above and the new parameter $a$ is an extra material constant.
Another model for shear-thinning is the Cross model \cite{Ro}
\begin{equation*}\label{eq4'}
\mu(\dot\gamma) =
\mu_\infty+\dfrac{\mu_0-\mu_\infty}{1+(\lambda\dot\gamma)^{(1-n)}}.
\end{equation*}
These models can also describe shear-thickening fluids, where $n>1$ and the meanings of $\mu_0, \mu_\infty$ exchange as follows
$$
\mathop {\lim }\limits_{\dot\gamma \to 0 }\mu(\dot\gamma)=\mu_\infty,\qquad  \mathop {\lim }\limits_{\dot\gamma\to \infty }\mu(\dot\gamma)=\mu_0.
$$
It is easy to see that the structural condition holds also for these models.

The viscosity of the original Bingham model is not a continuous function of the shear rate.
Such a discontinuous function is not suitable for numerical simulations \cite{Oh}. In \cite{Pap}, Papanastasiou proposed
the following model as a smoothed version of the Bingham model:
\begin{equation*}
\mu(\dot {\gamma})=\frac{\tau_0}{\dot {\gamma}}(1-e^{-m\dot
{\gamma}})+\eta_p.
\end{equation*}
Here $\tau_0$ is the yield stress, $m$ is the stress growth exponent (regularization parameter), and $\eta_p$ is the plastic
viscosity. Obviously, $\dot\gamma\mu(\dot\gamma)$ is a monotone increasing function of $\dot\gamma\geq 0$.

Another discontinuous model is the Casson model \cite{casson}. Its following smoothed version
\begin{equation*}
\mu(\dot\gamma)=\Big[\sqrt{\frac{\tau_0}{\dot
{\gamma}}}(1-e^{-\sqrt{m\dot {\gamma}}})+\sqrt{\eta_p}\Big]^2
\end{equation*}
was introduced in \cite{Neo}.
Here the parameters $\tau_0, m$ and $\eta_p$ are same as in the last model.
It is obvious that $\dot\gamma\mu(\dot\gamma)$ is a monotone increasing function of $\dot\gamma\geq 0$.

We conclude this section with the Powell-Eyring model \cite{Ro}
$$
\mu(\dot\gamma)=\mu_\infty+(\mu_0-\mu_\infty)\frac{\sinh^{-1}(\lambda\dot {\gamma})}{\lambda\dot {\gamma}},
$$
where $\mu_0,\mu_\infty$ and $\lambda$ are material constants, $
\mathop {\lim }\limits_{\dot {\gamma}\to 0 }\mu(\dot\gamma)=\mu_0$ and $\mathop {\lim }\limits_{\dot {\gamma} \to \infty }\mu(\dot\gamma)=\mu_\infty.
$
Obviously, the magnitude of the stress tensor $\dot\gamma\mu(\dot\gamma)$ for this model is also monotone increasing.

\section{Summary}
In this article we present a general methodology to conduct a detailed
asymptotic analysis of the two LB-BGK models for generalized
Newtonian fluids. We would like to point out that our analysis is quite
different from that based on the Chapman-Enskog expansion, which leads
to the compressible Navier-Stokes equations. Our analysis uses the Hilbert expansion and gives the incompressible
equations directly. It can expose certain important features of the LB solutions. As shown in
Section 3, such an analysis provides a theoretical basis not only for
using the Richardson extrapolation technique to improve the accuracy up
to higher orders, but also for using the iteration method to compute the
the rate-of-strain tensor. The latter makes sense specially for generalized
Newtonian fluids.

In contrast to the analysis for classical Newtonian fluids \cite{JY, JKL}, we have
introduced the lattice spacing $h$ into the equilibrium distributions in
\eqref{25} and \eqref{211}. More importantly, we have to deal with the
possibly non-zero terms $F^{(n)}$ in \eqref{310} and $A^{(n)}$ in \eqref{eq3'}
properly.

As a by-product of our analysis, a seemingly new structural condition on the
generalized Newtonian fluids is singled out. This condition is that
{\it the magnitude of the stress tensor increases with increasing the shear rate} (magnitude
of the rate-of-strain tensor). We verify this condition for all the existing constitutive
relations which are known to us. By inductive reasoning, we suggest to take this structural condition
as a criterion in constructing further constitutive relations for generalized Newtonian fluids.

Finally, let us mention that our analysis can be easily extended to the multiple-relaxation-time (MRT) models \cite{dH, LL, dhumieres02} or to three-dimensional lattices, although it is presented only for BGK models on the two-dimensional lattice D2Q9 \cite{QDL}.

\section*{Appendix}

In this Appendix we derive the equations in \eqref{311} and prove Theorem 1 for the two LB models defined in \eqref{25} and \eqref{211}.
For this purpose, we will often use, without notice, the
following simple facts that $w_i$ is even, $c_i$ is odd,
\begin{equation}\label{a1}
\sum_{i=0}^8w_i=1, \qquad \sum_iw_ic_i^tc_i=\frac{1}{3}I_2
\end{equation}
with $I_2$ the unit matrix of order 2. From these facts we easily
deduce that
\begin{equation}\label{a2}
\begin{array}{rl}
\sum_i f_{iL}(\rho, v)\equiv \rho, \qquad \sum_if_{iQ}(u, v)\equiv
0,\\[4mm]
\sum_ic_if_{iL}(\rho, v)\equiv v, \qquad \sum_ic_if_{iQ}(u, v)\equiv
0.
\end{array}
\end{equation}
The basic idea of our proofs is similar to that in \cite{JY, JKL}, but the possibly non-zero terms $F^{(n)}$ in \eqref{310} and $A^{(n)}$ in \eqref{eq3'} have to be treated properly.

\subsection*{A1.\quad The first model}
We begin with the hierarchy of equations in \eqref{38}--\eqref{310}:
\begin{align}
&h^1: \quad  f_i^{(1)}=f_{iL}(\rho^{(1)},v^{(1)}),\label{a3}\\
&h^2: \quad
\tau(S^{(1)})D_{i,1}f_i^{(1)}+f_i^{(2)}=f_{iL}(\rho^{(2)},v^{(2)})
+f_{iQ}(v^{(1)},v^{(1)}),\label{a4}\\
&h^k:\quad
\tau(S^{(1)})\sum_{s=1}^{k-1}D_{i,s}f_i^{(k-s)}+F^{(1)}\sum_{s=1}^{k-2}D_{i,s}f_i^{(k-1-s)}+\cdot\cdot\cdot+
F^{(k-2)}D_{i,1}f_i^{(1)}+f_i^{(k)}\nonumber\\
&=f_{iL}(\rho^{(k)},v^{(k)})+\sum_{p+q=k}f_{iQ}(v^{(p)},v^{(q)})\label{a5}
\end{align}
for $k\geq 3$. Summing up two sides of $\eqref{a4}$ over $i$ and
using \eqref{a2} together with $D_{i,1}=c_i\cdot\nabla$, we obtain
$$
0=\tau(S^{(1)})\sum_ic_i\cdot \nabla f_i^{(1)}
=\tau(S^{(1)})\nabla\cdot v^{(1)}.
$$
Here we have used the definitions of $\rho^{(2)}$ and $v^{(1)}$
given in \eqref{32}. Thus we have
\begin{equation}\label{a6}
\nabla\cdot v^{(1)}=0
\end{equation}
for $\tau(S^{(1)})=3\mu(S^{(1)}:S^{(1)}) +\frac{1}{2}>0$. Secondly, we multiply
$\eqref{a4}$ with $c_i$ and sum up the resultant equality to obtain
$$
\tau({S^{(1)}})\sum_ic_i(c_i\cdot \bigtriangledown f_i^{(1)})=0 .
$$
Moreover, it follows from \eqref{a1} and the expression of
$f_i^{(1)}$ given in \eqref{a3} that
\begin{equation*}\label{eq5}
\nabla\rho^{(1)}=0 .
\end{equation*}
Thus, $\rho^{(1)}$ is spatially homogenous.

Similarly, we deduce from \eqref{a5} with $k=3$ that
$$
\begin{array}{rl}
0=& \tau({S^{(1)}})(\sum_iD_{i,2}f_i^{(1)}+\sum_ic_i\cdot\nabla
f_i^{(2)})+ F^{(1)}\sum_ic_i\cdot\nabla
f_i^{(1)}\\[4mm]
= & \tau({S^{(1)}})(\sum_iD_{i,2}f_i^{(1)}+\sum_ic_i\cdot\nabla
f_i^{(2)}),\\[4mm]
 0=& \tau({S^{(1)}})(\sum_ic_i D_{i,2}f_i^{(1)}+
\sum_ic_ic_i\cdot\nabla f_i^{(2)})+ F^{(1)}\sum_ic_i
c_i\cdot\nabla f_i^{(1)}\\[4mm]
=& \tau({S^{(1)}})(\sum_ic_i D_{i,2}f_i^{(1)}+ \sum_ic_i
c_i\cdot\nabla f_i^{(2)}).
\end{array}
$$
Thus, we have
\begin{equation}\label{a8}
\begin{array}{rl}
\sum_iD_{i,2}f_i^{(1)}+\sum_ic_i\cdot\nabla f_i^{(2)}=0,\\[4mm]
\sum_ic_i D_{i,2}f_i^{(1)}+ \sum_ic_i c_i\cdot\nabla f_i^{(2)}=0,
\end{array}
\end{equation}
for $\tau(S^{(1)})>0$. Note that $D_{i,2}=\partial_t +
(c_i\cdot\nabla)^2/2$ and $\rho^{(1)}$ is spatially homogenous. The
first equation in \eqref{a8} leads immediately to
\begin{equation*}\label{eq6}
\partial_t\rho^{(1)} + \nabla\cdot v^{(2)}=0.
\end{equation*}
This, together with the initial condition $\rho^{(1)}\mid_{t=0}=0$
and the periodicity of the problem, gives
\begin{equation*}
\rho^{(1)}=0,\qquad \nabla\cdot v^{(2)}=0.
\end{equation*}
Moreover, we use $\nabla\cdot v^{(1)}=0$ to compute the two parts in the second equation in
\eqref{a8}:
\begin{subequations}\label{a10}
\begin{align}
\sum_ic_i D_{i,2}f_i^{(1)}&=\dfrac{\partial v^{(1)}}{\partial t} +
\dfrac{1}{2}\sum_ic_i (c_i\cdot\nabla)^2f_{iL}(\rho^{(1)},v^{(1)}) =\dfrac{\partial
v^{(1)}}{\partial t}+
\dfrac{1}{6}\Delta v^{(1)}, \\[4mm]
\sum_ic_i(c_i\cdot\nabla)f_i^{(2)}&=\sum_ic_i(c_i\cdot\nabla)[f_{iL}(\rho^{(2)},v^{(2)})
+ f_{iQ}(v^{(1)},v^{(1)}) - \tau(S^{(1)})c_i\cdot\nabla
f_i^{(1)}],\nonumber\\[4mm]
& = \dfrac{\nabla\rho^{(2)}}{3}
+v^{(1)}\cdot\nabla
v^{(1)}-\nabla\cdot\dfrac{\tau(S^{(1)})}{3}S^{(1)}.
\end{align}
\end{subequations}
Here we have used the following identity
$$
\sum_iw_ic_i(c_i\cdot\nabla)(c_i\cdot u)(c_i\cdot v) = \frac{1}{9}
\begin{pmatrix}2\partial_x(u_1v_1) + \partial_x(u\cdot v) + \partial_y(u_2v_1 + u_1v_2)\\
2\partial_y(u_2v_2) + \partial_y(u\cdot v) + \partial_x(u_2v_1 + u_1v_2)\end{pmatrix}
$$
for $u=(u_1, u_2)$ and $v=(v_1, v_2)$.
By combining \eqref{a6} and \eqref{a10}, we arrive at \eqref{311}.

Furthermore, we can consecutively deduce from \eqref{a5} that
\begin{equation}\label{a11}
\sum_i\sum_{s=1}^{k-1}D_{i,s}f_i^{(k-s)}=0, \qquad
\sum_i\sum_{s=1}^{k-1}c_iD_{i,s}f_i^{(k-s)}=0
\end{equation}
for $k\geq 3$. Observe that $D_{\bar i, s}=(-1)^sD_{i, s}$.

Now we prove Theorem $\ref{eq13}$ by induction on $k$. For $k=0$,
the conclusion follows simply from the choice of $f_i^{(k)}=w_i$ and the evenness of $w_i$.

Assume $f_i^{(l)}=(-1)^lf_{\bar i}^{(l)}$ for $l\leq k$. We show
$f_i^{(k+1)}=(-1)^{k+1}f_{\bar i}^{(k+1)}$. The inductive assumption
simply implies that $v^{(l)}=0$ and thereby $S^{(l)}=0$ for even
$l\leq k$. With this fact, we observe that the quadratic term in
\eqref{a5} is always even and vanishes for $k$ odd. Indeed, for $k$ odd
we see from $p + q = k$ that one of $p$ and $q$ must be even and thereby the
corresponding $v^{(p)}$ or $v^{(q)}$ vanishes. Moreover, we recall that
$I_h =tr(S_h^2)/h^2$ and compute the coefficient
$$
F^{(m)}=\dfrac{1}{m!}\sum_{j=1}^m\dfrac{d^j\tau(I_h)}{dI^j}\Big|_{h=0}\sum\limits_{k_1
+ \cdots +
k_j=m}\dfrac{d^{k_1}I_h}{dh^{k_1}}\big|_{h=0}\dfrac{d^{k_2}I_h}{dh^{k_2}}\big|_{h=0}
\cdots \dfrac{d^{k_j}I_h}{dh^{k_j}}\big|_{h=0}
$$
by the chain rule, which can be proved by induction on $m$. From this formula and the expression of $I_h$ it is easy to see that
$$
F^{(m)}=0
$$
for all odd $m < k$ and
\begin{equation}\label{a12}
F^{(k)}=2\dfrac{d\tau(I_h)}{dI}\mid_{h=0}S^{(1)}:S^{(k+1)}\\
=6\dfrac{d\mu(I_h)}{dI}\mid_{h=0}S^{(1)}:S^{(k+1)}.
\end{equation}
for odd $k$.
Thus, we deduce from \eqref{a5} that
\begin{equation}\label{a13}
\begin{array}{rl}
f_i^{(k+1)}- & f_{iL}(\rho^{(k+1)}, v^{(k+1)})\equiv
g_i^{k}=(-1)^{k+1}g_{\bar i}^{k},\\[4mm]
f_i^{(k+2)} - & f_{iL}(\rho^{(k+2)}, v^{(k+2)}) - 2f_{iQ}(v^{(1)},
v^{(k+1)}) \\[4mm]
& + \tau(S^{(1)}) D_{i, 1}f_i^{(k+1)} + F^{(k)}D_{i,
1}f_i^{(1)} \equiv \tilde g_i^{k}=(-1)^{k}\tilde g_{\bar i}^{k}.
\end{array}
\end{equation}
Consequently, it suffices to prove that $\rho^{(k+1)}=0$ in case $k$
is even or $v^{(k+1)}=0$ in case $k$ is odd, for $f_{iL}(\rho, v) =
w_i(\rho+ 3c_i\cdot v)$.

For $k$ even, it follows from \eqref{a11}, the
inductive assumption and the oddness of $g_i^{k}$ defined in \eqref{a13} that
\begin{equation*}\label{512}
\begin{array}{rl}
0=& \sum_i\sum_{s=1}^{k+1}c_iD_{i,s}f_i^{(k+2-s)}=\sum_ic_iD_{i,1}f_i^{(k+1)} = \sum_ic_i(c_i\cdot\nabla)[f_{iL}(\rho^{(k+1)}, v^{(k+1)}) + g_i^{k}]\\[4mm]
=&
\sum_ic_i(c_i\cdot\nabla)f_{iL}(\rho^{(k+1)}, v^{(k+1)})=\nabla\dfrac{\rho^{(k+1)}}{3}.
\end{array}
\end{equation*}
Namely, $\rho^{(k+1)}$ is spatially homogeneous. Moreover, we deduce from \eqref{a11} that
$$
0=\sum_i\sum_{s=1}^{k+2}D_{i,s}f_i^{(k+3-s)}=\sum_iD_{i,1}f_i^{(k+2)}
+\sum_iD_{i,2}f_i^{(k+1)} =\partial_t\rho^{(k+1)} + \nabla\cdot
v^{(k+2)}.
$$
This, together with the initial condition $\rho^{(k+1)}\mid_{t=0}=0$
and the periodicity of the problem, gives $\rho^{(k+1)}=0$  and thereby
$\nabla\cdot v^{(k+2)}=0$.

For $k$ odd, $(k-1)$ is even and we have $\nabla\cdot v^{(k+1)} =0$ as above.
Moreover, we see from \eqref{a11} and \eqref{a13} that
$$
\begin{array}{rl}
0 = & \sum_ic_iD_{i,2}f_i^{(k+1)}+\sum_ic_iD_{i,1}f_i^{(k+2)} \\[4mm]
= &  \sum_ic_iD_{i,2}[f_{iL}(\rho^{(k+1)}, v^{(k+1)}) + g_i^{k}]
+ \sum_ic_iD_{i,1}[f_{iL}(\rho^{(k+2)},
v^{(k+2)}) \\[4mm]
& + 2f_{iQ}(v^{(1)}, v^{(k+1)}) - \tau(S^{(1)})D_{i,
1}f_i^{(k+1)} - F^{(k)}D_{i,
1}f_i^{(1)} + \tilde g_i^{k}]\\[4mm]
=& \sum_ic_iD_{i,2}f_{iL}(\rho^{(k+1)}, v^{(k+1)})
+ \sum_ic_iD_{i,1}[f_{iL}(\rho^{(k+2)},
v^{(k+2)}) \\[4mm]
& + 2f_{iQ}(v^{(1)}, v^{(k+1)}) - \tau(S^{(1)})D_{i,
1}f_i^{(k+1)} - F^{(k)}D_{i,
1}f_i^{(1)}]\\[4mm]
= & \dfrac{\partial v^{(k+1)}}{\partial
t} + \frac{1}{6}\Delta v^{(k+1)}
+ \frac{1}{3}\nabla \rho^{(k+2)} + v^{(k+1)}\cdot\nabla v^{(1)} +
v^{(1)}\cdot\nabla v^{(k + 1)} \\[4mm]
& - \frac{1}{3}\nabla\cdot[\tau(S^{(1)})S^{(k+1)} + F^{(k)}S^{(1)}].
\end{array}
$$
Here the computations in \eqref{a10} have been used in the last step. Then we have
\begin{equation}\label{a14}
\begin{array}{rl}
& \nabla\cdot v^{(k+1)}= 0, \\[4mm]
& \dfrac{\partial v^{(k+1)}}{\partial
t} +\nabla\dfrac{\rho^{(k+2)}}{3} +v^{(1)}\cdot\nabla
v^{(k+1)}+v^{(k+1)}\cdot\nabla
v^{(1)}\\[4mm]
= &
\dfrac{1}{3}\nabla\cdot\Big[\big(\tau(S^{(1)})-\dfrac{1}{2}\big)S^{(k+1)}\Big]
+\dfrac{1}{3}\nabla\cdot[F^{(k)}S^{(1)}].
\end{array}
\end{equation}

Next we show that $v^{(k+1)}=0$ by using the equations in \eqref{a14} together with the initial condition $v^{(k+1)}|_{t=0}=0$. Note that the equations in \eqref{a14} with $k$ odd are linear with respect to $v^{(k+1)}$. Taking the inner
product of the second equation in \eqref{a14} with $v^{(k+1)}$ and
integrating the resultant equality, we get
\begin{equation*}
\begin{array}{rl}
& \dfrac{1}{2}\dfrac{\partial}{\partial
t}\int_{\Omega}v^{(k+1)}\cdot v^{(k+1)}dx +
\int_{\Omega}v^{(k+1)}\cdot\nabla\dfrac{\rho^{(k+2)}}{3}dx\\[4mm]
+ & \int_\Omega v^{(k+1)}\cdot(v^{(1)}\cdot\nabla) v^{(k+1)}dx
+\int_\Omega v^{(k+1)}\cdot(v^{(k+1)}\cdot\nabla) v^{(1)}dx\\[4mm]
=& \dfrac{1}{3}\int_{\Omega}v^{(k+1)}\cdot\nabla\cdot
\Big[\big(\tau(S^{(1)})-\frac{1}{2}\big)S^{(k+1)}\Big] +
\dfrac{1}{3}\int_{\Omega}v^{(k+1)}\cdot\nabla\cdot[F^{(k)}S^{(1)}].
\end{array}
\end{equation*}
Set $\|u\|_{L^2}^2:=\int_{\Omega}u\cdot udx$,
$\|u\|_{L^\infty}:=\sup_{x\in\Omega}\sqrt{u\cdot u}$ and $I^{(1)}=tr(S^{(1)}:S^{(1)})$. Integrating by parts and
using $\nabla\cdot v^{(1)}=\nabla\cdot v^{(k+1)}=0$, we deduce from the last
equation and $F^{(k)}=6\mu'(I^{(1)})S^{(1)}:S^{(k+1)}$ in \eqref{a12} that
\begin{equation*}
\begin{array}{rl}
& \dfrac{1}{2}\dfrac{d}{dt}\|v^{(k+1)}\|_{L^2}^2\\[4mm]
= & -\int_\Omega v^{(k+1)}\cdot(v^{(k+1)}\cdot\nabla) v^{(1)}dx
-\int_\Omega \mu(I^{(1)})S^{(k+1)}:\nabla v^{(k+1)}dx
-\dfrac{1}{3}\int_\Omega F^{(k)}S^{(1)}:\nabla v^{(k+1)}dx \\[4mm]
\leq & \|\nabla v^{(1)}\|_{L^\infty}\|v^{(k+1)}\|_{L^2}^2
- \dfrac{1}{2}\int_\Omega \mu(I^{(1)})S^{(k+1)}:S^{(k+1)}dx
-\dfrac{1}{6}\int_\Omega F^{(k)}S^{(1)}:S^{(k+1)}dx\\[4mm]
= & \|\nabla v^{(1)}\|_{L^\infty}\|v^{(k+1)}\|_{L^2}^2 -
\dfrac{1}{2}\int_\Omega \mu(I^{(1)})S^{(k+1)}:S^{(k+1)}dx
-\int_\Omega\mu'(I^{(1)})(S^{(1)}:S^{(k+1)})^2dx\\[4mm]
\leq & \|\nabla v^{(1)}\|_{L^\infty}\|
v^{(k+1)}\|_{L^2}^2-\dfrac{1}{2}\int_\Omega[\mu(I^{(1)})+2\min\{\mu'(I^{(1)}),0\}
I^{(1)}]S^{(k+1)}:S^{(k+1)}dx .
\end{array}
\end{equation*}
Here the last inequality is due to the Cauchy-Schwartz inequality.
Thanks to $\mu(I)+2\min\{\mu'(I),0\}I\geq0$, the last estimate leads to
$$
\dfrac{d}{dt}\|v^{(k+1)}\|_{L^2}^2\leq 2\|\nabla v^{(1)}\|_{L^\infty}\|
v^{(k+1)}\|_{L^2}^2.
$$
This differential inequality together with the initial condition $v^{(k+1)}|_{t=0}=0$
simply gives $v^{(k+1)}=0$.

This completes the proof of Theorem 1 for the first model.

\subsection*{A2.\quad The second model}
This subsection is devoted to the second model. By equating the coefficient of $h^k$ in the two sides of \eqref{eq3'} and using $v^{(0)}=0$, we obtain the following equations
\begin{align}
&h^0:\quad f_{i}^{(0)}=f_{iL}(\rho^{(0)},v^{(0)})=f_{iL}(1,0),\nonumber\\
&h^1:\quad f_{i}^{(1)}=f_{iL}(\rho^{(1)},v^{(1)}),\label{b1}\\
&h^2:\quad \tau D_{i,1}f_i^{(1)}+f_i^{(2)}=f_{iL}(\rho^{(2)},v^{(2)})
+f_{iQ}(v^{(1)},v^{(1)})+w_iA^{(0)}S^{(1)}:c_i^tc_i, \label{b2}\\
&h^k:\quad
\tau(\sum_{s=1}^{k-1}D_{i,s}f_i^{(k-s)})+f_i^{(k)}
=f_{iL}(\rho^{(k)},v^{(k)})+\sum_{p+q=k}f_{iQ}(v^{(p)},v^{(q)})\nonumber\\
&+\sum_{a+b=k-1}w_iA^{(a)}S^{(b)}:c_i^tc_i \label{b3}
\end{align}
for $k\geq3$. Note that
\begin{equation}\label{b4}
\sum_iw_iA^{(a)}S^{(b)}:c_i^tc_i=A^{(a)}S^{(b)}:\sum_iw_ic_i^tc_i=\frac{2}{3}A^{(a)}\nabla\cdot v^{(b)}.
\end{equation}

Summing up the two sides of \eqref{b2} over $i$, we get
$$
\tau\sum_ic_i\cdot \nabla f_i^{(1)}=\frac{2}{3}A^{(0)}\nabla\cdot v^{(1)}
$$
and thereby
$$
(\tau-\frac{2}{3}A^{(0)})\nabla\cdot
v^{(1)}=0.
$$
This gives
\begin{equation}\label{b5}
\nabla\cdot v^{(1)}=0
\end{equation}
for $A^{(0)}=A(S^{(1)})$ and $\tau-\frac{2}{3}A(S^{(1)}) = \dfrac{1}{2} + 3\mu(S^{(1)}:S^{(1)})>0$.
On the other hand, we multiply the two sides of \eqref{b2} by $c_i$ and
sum up the resultant equalities over $i$ to obtain
$$
\tau\sum_ic_i(c_i\cdot \nabla f_i^{(1)})=0
$$
and thereby
\begin{equation*}\label{b6}
\nabla\rho^{(1)}=0 .
\end{equation*}
Thus, $\rho^{(1)}$ is spatially homogenous.

Similarly, we deduce from \eqref{b3} with $k=3$ that
\begin{equation*}
\begin{split}
\tau(\sum_iD_{i,2}f_i^{(1)}+\sum_iD_{i,1}f_i^{(2)})&=\sum_iw_iA^{(0)}S^{(2)}:c_i^tc_i +\sum_iw_iA^{(1)}S^{(1)}:c_i^tc_i\\
&=\frac{2}{3}A^{(0)}\nabla\cdot v^{(2)}+\frac{2}{3}A^{(1)}\nabla\cdot v^{(1)}.
\end{split}
\end{equation*}
Since $\nabla\cdot v^{(1)}=0$ and $\rho^{(1)}$ is spatially homogenous,
the last equation leads to
\begin{equation}\label{b0}
\dfrac{\partial\rho^{(1)}}{\partial t}+\nabla\cdot v^{(2)}=\frac{2A^{(0)}}{3\tau}\nabla\cdot v^{(2)}
\end{equation}
and thereby
\begin{equation*}
\Big(1- \frac{2A^{(0)}}{3\tau}\Big)^{-1}\dfrac{\partial}{\partial t}\rho^{(1)}=\nabla\cdot v^{(2)}
\end{equation*}
for $1-\frac{2}{3\tau}A>0$ and $A^{(0)}=A(S^{(1)})$. Integrating the last equation over the periodic domain and using the spatial homogeneousness of $\rho^{(1)}$ gives
$$
\dfrac{\partial\rho^{(1)}}{\partial t}\int\Big(1- \frac{2A^{(0)}}{3\tau}\Big)^{-1}dx=0.
$$
and thereby $\dfrac{\partial\rho^{(1)}}{\partial t}=0$. This together with the initial condition $\rho^{(1)}|_{t=0}=0$ implies that
\begin{equation}\label{b6}
\rho^{(1)}=0, \qquad \nabla\cdot v^{(2)}=0.
\end{equation}
On the other hand, we multiply the two sides of \eqref{b3} with $k=3$ by $c_i$ and
sum up the resultant equalities over $i$ to obtain
$$
\tau(\sum_ic_iD_{i,2}f_i^{(1)}+\sum_ic_iD_{i,1}f_i^{(2)})=0.
$$
From this, we deduce as from \eqref{a8} to \eqref{a10} that
\begin{subequations}\label{b7}
\begin{align}
\sum_ic_iD_{i,2}f_i^{(1)}&=\dfrac{\partial v^{(1)}}{\partial t}+\dfrac{1}{6}\bigtriangleup v^{(1)},\\[4mm]
\sum_ic_iD_{i,1}f_i^{(2)}&=\dfrac{\nabla\rho^{(2)}}{3}+v^{(1)}\cdot\nabla v^{(1)}-\frac{\tau}{3}\triangle v^{(1)}+\frac{2}{9}\nabla\cdot(A^{(0)}S^{(1)})
\end{align}
\end{subequations}
By combining these with \eqref{b5}, we arrive at \eqref{311}.

Now we turn to prove Theorem 1 for the second model again by induction on $k$.
For $k=0$, the conclusion follows simply from the choice of $f_i^{(k)}=w_i$.

Assume $f_i^{(l)}=(-1)^lf_{\bar i}^{(l)}$ for $l\leq k$. We show
$f_i^{(k+1)}=(-1)^{k+1}f_{\bar i}^{(k+1)}$. As in the previous subsection, we deduce from \eqref{b3} that
\begin{equation}\label{b9}
\begin{array}{rl}
f_i^{(k+1)}- & f_{iL}(\rho^{(k+1)}, v^{(k+1)})\equiv
g_i^{k}=(-1)^{k+1}g_{\bar i}^{k},\\[4mm]
f_i^{(k+2)} - & f_{iL}(\rho^{(k+2)}, v^{(k+2)}) - 2f_{iQ}(v^{(1)},
v^{(k+1)}) \\[4mm]
& + \tau D_{i, 1}f_i^{(k+1)} - w_iA^{(0)}S^{(k+1)}:c_i^tc_i - w_iA^{(k)}S^{(1)}:c_i^tc_i\equiv \tilde g_i^{k}=(-1)^{k}\tilde g_{\bar i}^{k}.
\end{array}
\end{equation}
Consequently, it suffices to prove that $\rho^{(k+1)}=0$ in case $k$
is even or $v^{(k+1)}=0$ in case $k$ is odd, for $f_{iL}(\rho, v) =
w_i(\rho+ 3c_i\cdot v)$.

For $k$ even, it follows from \eqref{b3}, the
inductive assumption and the oddness of $g_i^{k}$ defined in \eqref{b9} that
\begin{equation*}\label{512}
\begin{array}{rl}
0=& \sum_i\sum_{s=1}^{k+1}c_iD_{i,s}f_i^{(k+2-s)}=\sum_ic_iD_{i,1}f_i^{(k+1)} = \sum_ic_i(c_i\cdot\nabla)[f_{iL}(\rho^{(k+1)}, v^{(k+1)}) + g_i^{k}]\\[4mm]
=&
\sum_ic_i(c_i\cdot\nabla)f_{iL}(\rho^{(k+1)}, v^{(k+1)})=\nabla\dfrac{\rho^{(k+1)}}{3}.
\end{array}
\end{equation*}
Namely, $\rho^{(k+1)}$ is spatially homogeneous. Moreover, we deduce from \eqref{b3} and \eqref{b4} that
$$
\frac{2}{3\tau}A^{(0)}\nabla\cdot v^{(k+2)}=\sum_i\sum_{s=1}^{k+2}D_{i,s}f_i^{(k+3-s)}=\sum_iD_{i,1}f_i^{(k+2)}
+\sum_iD_{i,2}f_i^{(k+1)} =\partial_t\rho^{(k+1)} + \nabla\cdot
v^{(k+2)}.
$$
As from \eqref{b0} to \eqref{b6}, this  gives $\rho^{(k+1)}=0$ and thereby
$\nabla\cdot v^{(k+2)}=0$.

For $k$ odd, $(k-1)$ is even and we have $\nabla\cdot v^{(k+1)} =0$ as above.
Moreover, we see from \eqref{b3} and \eqref{b9} that
$$
\begin{array}{rl}
0 = & \sum_ic_iD_{i,2}f_i^{(k+1)}+\sum_ic_iD_{i,1}f_i^{(k+2)} \\[4mm]
= &  \sum_ic_iD_{i,2}[f_{iL}(\rho^{(k+1)}, v^{(k+1)}) + g_i^{k}]
+ \sum_ic_iD_{i,1}[f_{iL}(\rho^{(k+2)}, v^{(k+2)}) \\[4mm]
& + 2f_{iQ}(v^{(1)}, v^{(k+1)}) - \tau D_{i,1}f_i^{(k+1)} + w_iA^{(0)}S^{(k+1)}:c_i^tc_i + \tilde g_i^{k}]\\[4mm]
=& \sum_ic_iD_{i,2}f_{iL}(\rho^{(k+1)}, v^{(k+1)})
+ \sum_ic_iD_{i,1}[f_{iL}(\rho^{(k+2)}, v^{(k+2)}) \\[4mm]
& + 2f_{iQ}(v^{(1)}, v^{(k+1)}) - \tau D_{i, 1}f_i^{(k+1)} + w_iA^{(0)}S^{(k+1)}:c_i^tc_i +w_iA^{(k)}S^{(1)}:c_i^tc_i]\\[4mm]
= & \dfrac{\partial v^{(k+1)}}{\partial
t} + \dfrac{1}{6}\Delta v^{(k+1)}
+ \dfrac{1}{3}\nabla \rho^{(k+2)} + v^{(k+1)}\cdot\nabla v^{(1)} +
v^{(1)}\cdot\nabla v^{(k + 1)} \\[4mm]
& - \dfrac{\tau}{3}\Delta v^{(k+1)} + \dfrac{2}{9}\nabla\cdot [A^{(0)}S^{(k+1)}] + \dfrac{2}{9}\nabla\cdot [A^{(k)}S^{(1)}].
\end{array}
$$
Here the computations in \eqref{b7} have been used in the last step.
Thus we have
\begin{equation}\label{b10}
\begin{array}{rl}
& \nabla\cdot v^{(k+1)}= 0, \\[4mm]
& \dfrac{\partial v^{(k+1)}}{\partial
t} +\nabla\dfrac{\rho^{(k+2)}}{3} +v^{(1)}\cdot\nabla
v^{(k+1)}+v^{(k+1)}\cdot\nabla
v^{(1)}\\[4mm]
= &
\nabla\cdot[\mu(S^{(1)}:S^{(1)})S^{(k+1)}] -
\dfrac{2}{9}\nabla\cdot [A^{(k)}S^{(1)}],
\end{array}
\end{equation}
for
$$
\dfrac{1}{3}(\tau-\dfrac{1}{2})-\dfrac{2}{9}A^{(0)}=\mu(S^{(1)}:S^{(1)}).
$$
Since
$$
A^{(k)}=-9\mu'(S^{(1)}:S^{(1)})S^{(1)}:S^{(k+1)}
$$
which is similar to $F^{(k)}$ in \eqref{a12}, the equations in \eqref{b10} is exactly same as those in \eqref{a14} and, hence, we have $v^{(k+1)}= 0$.

This completes the proof.

\section*{Acknowledgments}

This work was supported by the Tsinghua University Initiative Scientific Research Program (20121087902) and by Specialized Research Fund for the Doctoral Program of Higher Education (Grant No. 20100002110085).

\end{spacing}
\end{document}